\documentclass[%
 aip,jmp,amsmath,amssymb,reprint
]{revtex4-2}

\usepackage{datetime}
\usepackage[version=4]{mhchem}
\usepackage{times,mathptmx}
\usepackage{graphicx}
\usepackage{wrapfig}
\usepackage{dcolumn}
\usepackage{eucal}
\usepackage{relsize}
\usepackage{physics}
\usepackage{sidecap}

\def\ie{{\it i.e.},~}

\def\r{\right}
\def\l{\left}

\def\2D{\mathsf{2D}}

\begin{document}
\title[working draft]{
What do the fast dynamics tell us about aggregation?
}

\author{Tamoghna Das}
\email{tamoghna.4119@gmail.com}
\affiliation{%
Center for Soft and Living Matter,\\ Institute for Basic Science (IBS), Ulsan,\\ 44919, Republic of Korea
}%
 

\date{\today}

\begin{abstract}
Two typical morphology of two-dimensional aggregates are considered: compact crystalline clusters and string-like non-compact conformations. Simulated trajectories of both types of aggregates are analysed with fine spatial resolution. While the long-time geometry of such trajectories appears to be statistically identical for two conformations, the self-overlap statistics reveal two distinct short-time {\em pre-caging} mechanisms. While the time-scale is directly proportional with length-scale for particles in a compact aggregates, a inverse relationship holds for non-compact clusters. These short length-fast time relationship of particle localization might hold the key to the structure-function relationship of aggregate forming systems and other non-equilibrium soft materials.
\end{abstract}


\maketitle


{\bf Introduction --} Competition between the length and time scales lies at the heart of the formation of self-assembled structures.~\cite{Penrose:1966, Stell:1970, Amar:1991} This competition happens most naturally in presence of a short-range attraction and a long-range repulsion as it appears in systems of astronomical length scale, such as pasta phase in neutron stars~\cite{Piekarewicz:2004} to nano-scale metallic, organometallic and polymeric self-assemblies to biologically relevant macro-molecule formations. While the exact nature of the interactions vary widely depending on the systems, geometric frustration originating from the competing interactions leads to the localization or {\em caging} of the constituent units. The emergent finite length and time scales of {\em the cage} then governs the global structure and dynamics of the resulting self-assembled structures. Over the past decades, this picture has been actively constructed and revised through numerous theoretical and experimental efforts to settle down to its current state. The local dynamics before caging has attracted very little attention though as it is considered to be of ballistic nature. In this study, we show that the effect of geometric frustration goes to even smaller length and time scales, deep inside the ballistic regime.

Specifically, we have analysed the simulated trajectories of an aggregate forming system in two dimensions. Two limiting aggregate conformations are chosen for comparison: {\em compact} clusters with local crystalline arrangement of particles and {\em string-like} non-compact aggregates. Although these are the steady state structures, \ie{} no coarsening is observed, they are highly dynamic. Particles detaches from their parent clusters and move freely before they reattach to the same or a neighboring cluster. We refer to this intermediate excursion of particles, in between their cages, as the {\em pre-caging} dynamics. Examining the self-overlap of a particle as a function of certain bounding length scale informs us about this entire dynamic behaviour. As it turns out that this intermediate cage-free motion is not ideally free as it must be under the influence of the underlying potential energy landscape. In fact, the nature of pre-caging in two different aggregation environments are found to be quite opposite which reveals certain intriguing relation between the fast dynamics and the energy landscape as explained later.

{\bf Model aggregate --} The particular realization of competing interaction used in this study is a superposition of a short-range attraction ($\Phi_{SA}$) and a long-range repulsion ($\Phi_{LR}$) of the following forms: 
$\Phi_{SA}=4\epsilon\l[(\sigma/r)^{2\alpha}-(\sigma/r)^{\alpha}\r]$ and $\Phi_{LR}=(A\sigma/r)\exp(-r/\xi)$ where $\sigma$ is the particle diameter. Setting the energy scale by $\epsilon$, the strength of attraction is matched with the repulsion strength $A=4\epsilon$. The range of attraction is fixed at $0.2\sigma$ by setting $\alpha=18$ and the range of repulsion $\xi$ is tuned to get different morphologies. While this realization closely mimics the polymer-grafted nano particle systems, it is by no means the most generic one. However, for any system with competing interactions, it is possible to define a single controlling parameter as the ratio of relevant length scales as shown in an earlier study.~\cite{Bandi:2015} We prepare our system starting from a high temperature ($T=\epsilon$) liquid configuration and cooling it very slowly to a temperature $T=0.05\epsilon$, well below the micro-phase separation boundary.~\cite{Reichman:2007} Temperature drops by $10^{-4}\epsilon$ per unit time $\tau=\sqrt{\sigma^2/\epsilon}$. Details of this procedure along with other long-time dynamical characterization are presented elsewhere.~\cite{Bandi:2016}

\begin{figure*}[h!]
\includegraphics[width=0.73\textwidth]{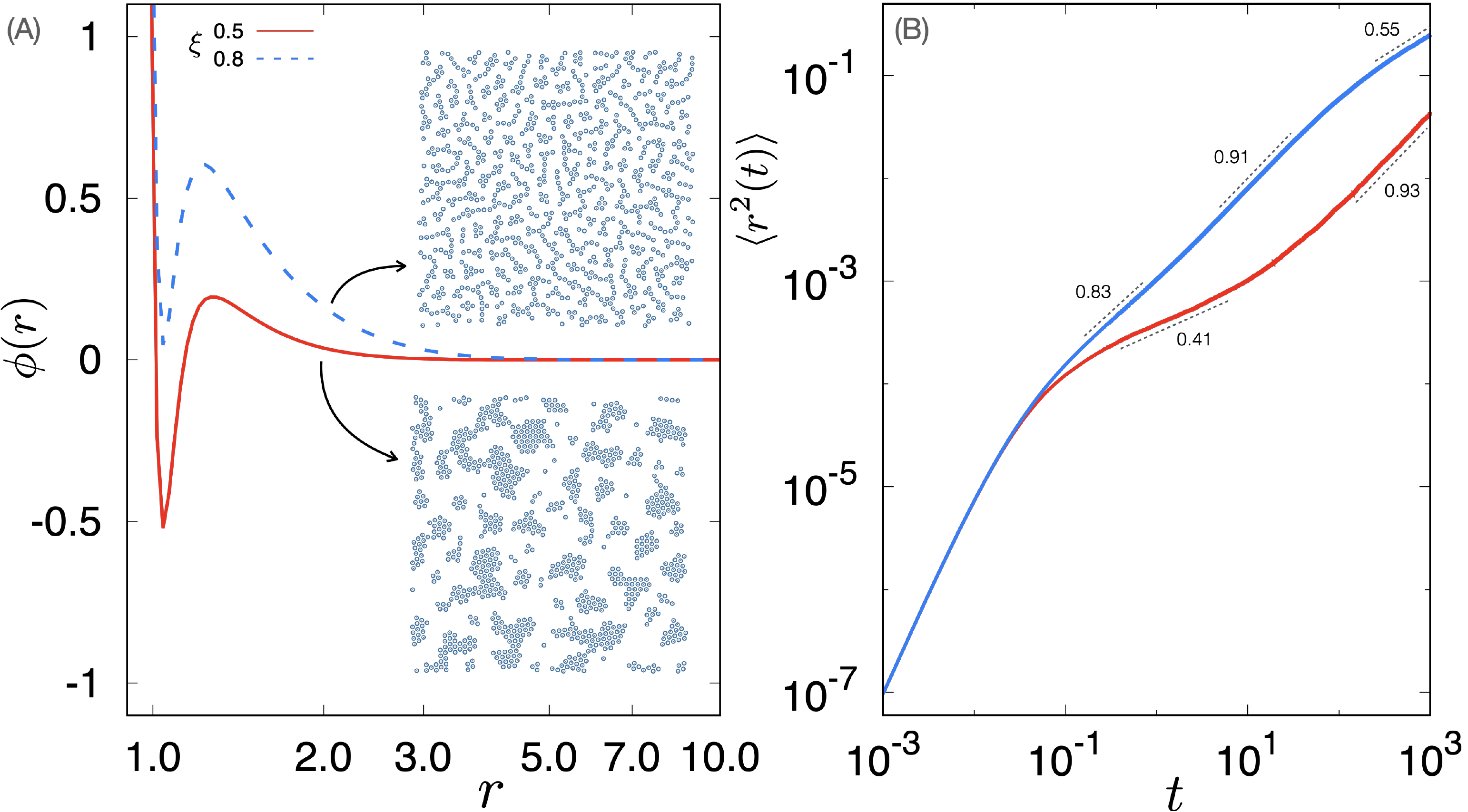}
\caption{\label{fig:trajMSD}
{\bf Global conformation and global dynamics:}
(A) The form of the potential $\phi(r)$ is plotted in a linear-log scale as a function of pair distance $r$. Note the presence of an energy barrier barrier which introduces a geometric frustration by separating the length scales. By tuning the height of the barrier, one can seamlessly achieve aggregates between compact and string-line conformations, which are shown in the {\em Inset}. (B) Mean square displacement (MSD) for both of the steady state limiting conformations shows multiple time step relaxation. Details are described in the text.
}
\end{figure*}

{\bf Results --} For the present study, we have chosen two specific values of $\xi$. Note the presence of the energy barrier in the effective potential, $\phi(r)=\phi_{SA}+\phi_{LR}$ as a result of the competition between the two length scales of the two contributing potentials.~(Fig.\ref{fig:trajMSD}A) This is the key to the geometric frustration experienced by a particle in a many-body configuration. By tuning $\xi$, the height of this barrier and the nature of the minimum behind this barrier can be controlled. At $\xi=0.5$, there is a negative global minimum in $\phi(r)$ and the particles, under the dominant attraction, aggregates into compact, locally crystalline arrangement of finite size.~(Fig.\ref{fig:trajMSD}A {\em Inset}) As the repulsive length scale is increased to $\xi=0.8$, the global $\phi(r)$ minimum becomes local and now, protected by an even stronger energy barrier. As a result, particles arrange themselves in ramified structures of nearly string-like conformations.~(Fig.\ref{fig:trajMSD}A {\em Inset}) Aside from these global structural changes, the geometric frustration affects the global relaxation dynamics of the system as well. Particles in the two systems under consideration show very different long-time dynamics as captured by their mean-squared displacement (MSD) $\langle r^2(t)\rangle$ as a function of time.~(Fig.\ref{fig:trajMSD}B) Compact cluster system shows a ballistic ($\langle r^2(t)\rangle \sim t^2$) to diffusive ($\langle r^2(t)\rangle \sim t$) crossover through a strong intermediate time relaxation ($\sim t^{0.4}$) spanned over almost two decades of time scale. This intermediate relaxation is much more gentle for the string-like aggregates ($\langle r^2(t)\rangle \sim t^{0.8}$) before it turns to diffusive behaviour. However, the diffusive dynamics does not persist in the long time, instead, it turns to a typical sub-diffusion ($\sim t^{0.5}$) within our observation time. We speculate that the ramification of available space due to the specific aggregation morphology in this system is responsible for such behaviour. However, we leave this aspect for future exploration and focus on the individual trajectories, instead, to understand the dynamics of the departure from ballistic behaviour.

\begin{figure*}[h!]
    \includegraphics[width=0.73\textwidth]{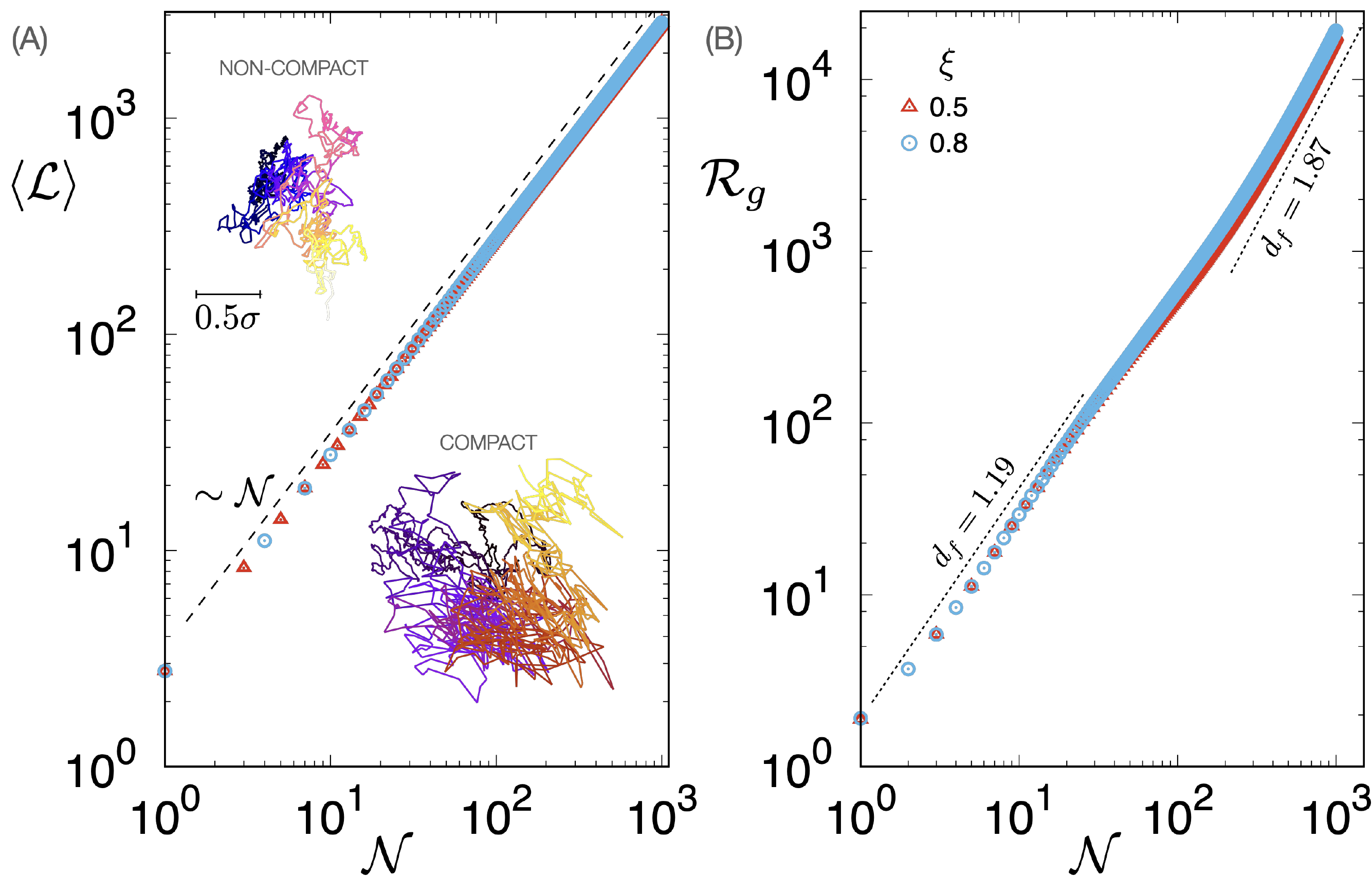}
\caption{\label{fig:trajGeom}
{\bf Statistical properties of the trajectories:} 
Envisioning the trajectories as growing polymers, average end to end distance $\langle\mathcal{L}\rangle$ of this polymers is plotted as function of the number of monomers $\mathcal{N}$. This shows an average linear growth of polymers (trajectories) as a function of increasing $\mathcal{N}$ (time) for both conformations. Note that individual trajectories in each conformations visually appear to be quite different as plotted in the {\em Inset}. Each representative trajectory is of length $10^3\tau$ and plotted with the same length scale as shown by a bar. (B) Radius of gyration $\mathcal{R}_g$ plotted as function of $\mathcal{N}$ points to a crossover timescale as the overall shape of the trajectories goes from linear to circular as identified by their respective fractal dimensions. However, this measure, again, does not distinguish between the global conformations.
}
\end{figure*}

The overall shape of representative trajectories from two different conformations appear to be very different.~(Fig.\ref{fig:trajGeom}A{\em Inset}) Particles in compact clusters appear to explore the space in such a way that over certain long time the trajectory fills a nearly circular region. In comparison, the trajectory geometry of particles in string-like aggregates appears to be more jagged in space as the particle moves within a narrow restricted space. The spatial localization and escape from that is more evident though in this latter case of string-like aggregates. We first investigate the statistical properties of these trajectories imagining them as a growing polymers. Each monomer in these polymers would represent the position of a particle at a certain time. Total number of monomers $\mathcal{N}$ would then be exactly equal to the total time $t$ over which the trajectories are recorded. Unlike the usual polymers, such trajectory polymers would not be self-avoiding and would be highly entangled in space. Yet, we find that both end-to-end distance $\langle\mathcal{L}\rangle$ and the radius of gyration $\mathcal{R}_g$ of these polymers (trajectories) are unable to pick up any such signature. $\langle\mathcal{L}\rangle$ grows linearly with $\mathcal{N}$ for trajectories in both of the conformations.~(Fig.\ref{fig:trajGeom}A) The shape of the polymer, quantified by $\mathcal{R}_g$ is related to the number of monomers, $\mathcal{N}\sim\mathcal{R}_g^{d_f}$, where $d_f$ is the fractal dimension of the polymer. For the trajectory polymers, a crossover from nearly linear ($d_f\sim 1$) to almost circular shape ($d_f\sim 2$) is observed as a function of increasing $\mathcal{N}$ (time).~(Fig.\ref{fig:trajGeom}B) We mention that the large-$\mathcal{N}$ behaviour is reported for localization of DNA into knots under strong adsorption.~\cite{Dietler:2007} Though the shape crossover corresponds to the departure from ballistic dynamics, these measures turn out to be inadequate to provide any further detailed information about the process and also unable to distinguish between the trajectories from two different morphologies. This prompts us to examine the finer details of trajectories with the aid of self-overlap function.

How long does a particle stay at a specific position is quantified with the self-overlap function by computing total number of occurrence of a particle within a small cut-off radius $\sigma_R$ around its initial position, over time $t$. Formally, the measure reads as $Q_s(t)=\sum_{j=1}^N \omega(|r_j(t)-r_j(0)|)$, with $\omega=1$ when $|r_j(t)-r_j(0)|<\sigma_R$ and $\omega=0$ otherwise. Typically, $\sigma_R$ is interpreted as the caging length and there is an active debate about how to assign a reasonable value to it in the context of glassy materials.~\cite{Stillinger:1985, Ma:2001, Glotzer:2003, Douglas:2019} Here, as we are only interested in the pre-caging dynamics, we adopt a much straight-forward definition. We choose the upper bound of $\sigma_R^M=\sqrt{\langle r^2(\tau_b)\rangle}=10^{-2}\sigma$ where $\tau_b$ is the ballistic timescale determined from MSD and study the behaviour of $Q_s(t)$ for a set of $\sigma_R\leq\sigma_R^M$.

\begin{figure*}[h!]
    \includegraphics[width=0.99\textwidth]{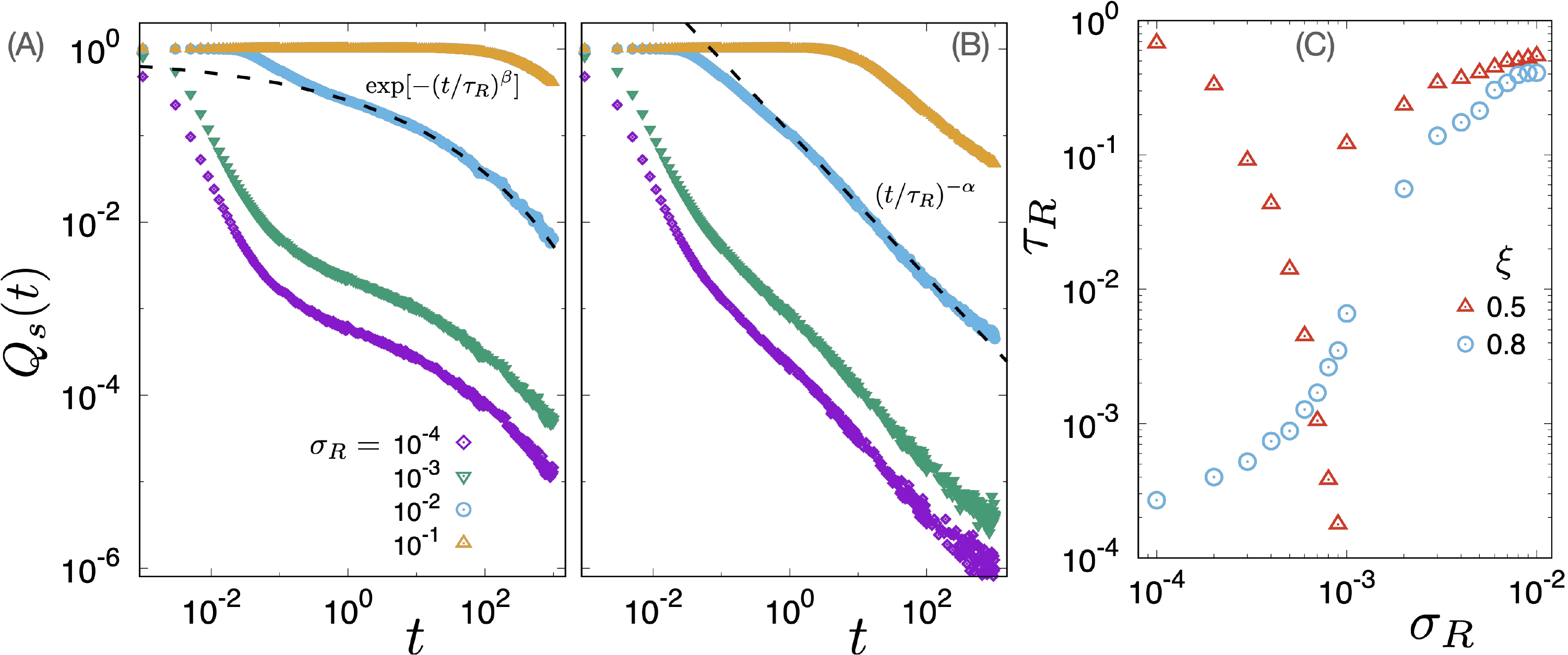}
\caption{\label{fig:selfOverlap}
{\bf Space-time relation from self-overlap of a particle:} Temporal behaviour of the probability of finding a particle within the specified length-scale $\sigma_R$ shows (A) stretched exponential decay for compact clusters and (B) power-law decay for non-compact clusters. The residence time $\tau_R$ extracted from these distributions are plotted in (C) as a function of corresponding residence length $\sigma_R$. This short scale space-time relationship clearly shows that the departure from the ballistic behaviour happens in opposite fashion in two different aggregate conformations as detailed in the text. 
}
\end{figure*}
We notice that for both of the aggregate conformations, the early time decay of $Q_s(t)$ is Gaussian but their long time decay behaviour is quite different. For compact clusters, this long time decay is best fitted with a stretched exponential,~(Fig.\ref{fig:selfOverlap}A) whereas, for string-like clusters, it decays algebraically.~(Fig.\ref{fig:selfOverlap}B) While the width of the early time peak of $Q_s(t)$ would inform us about the localization time within the specified $\sigma_R$, the departure of the particle from this localization zone and its occasional revisit would contribute to the long time decay of $Q_s(t)$. We could then estimate the residency time $\tau_R$ of a particle within $\sigma_R$ in a statistical sense by fitting this long time tail of $Q_s(t)$. We find that this space-time ($\sigma_R$-$\tau_R$) relationship is opposite within the two aggregation environments.~(Fig.\ref{fig:selfOverlap}C)

For the smallest $\sigma_R$ considered, the typical residence time $\tau_R$ for particles in string-like clusters is almost equal to one unit of time, $\tau$, which is surprisingly large. As we increase $\sigma_R$, up to a certain value, $\tau_R$ decreases monotonically by almost $4$-orders of magnitude before it jumps directly to a much larger value and slowly approaches $\tau_B$. We relate this curious localization behaviour with the specific form of pair potential for $\xi=0.8$ which has a positive local minimum behind a strong positive energy barrier. While any localization at the local minimum is strongly protected by the barrier, any deviation from that is very much unfavorable for the barrier and the discrete jump is indicative of the escape from the barrier. For particles in a compact cluster ($\xi=0.5$) are, in comparison, much more mobile as it has very small $\tau_R$ at a very small $\sigma_R$. In this case, the global negative minimum of the potential might be inaccessible due to the positive energy barrier. As with increasing $\sigma_R$, more of the accessible region can be probed and $\tau_R$ is found to be increasing monotonically. However, note the sharp rise in $\tau_R$ occurring at the same $\sigma_R$ as for the $\xi=0.8$ case. This supports our previous intuition that this sharp change is related to the crossing of the energy barrier. Following the same line, we would argue that $\sigma_R$, probed by $Q_s(t)$, reveals an effective reaction coordinate to study the spatio-temporal features of a vast class of aggregate forming systems.

{\bf Conclusion --} In essence, this study reveals a non-trivial relationship between the fast ballistic dynamics and the potential energy landscape of a non-equilibrium system which to our knowledge was not addressed before. Most interesting outcome of this project is to find out that local minimum in the effective interaction might lead to stronger localization than the global minimum for a system with competing interactions. The actual correspondence between the exact topology of the potential energy landscape and the probed reaction coordinate needs to be established in detail though. Nevertheless, this study provides us with a fresh perspective to study complex dynamical response in soft materials in much simpler settings, in terms of suitably constructed barrier crossing problems. Further investigations are also required to understand how this fast dynamics might (not) affect the slow dynamical modes of such system. Specifically, for the aggregate forming systems, the ballistic space-time correspondence uncovered here might be useful to develop a microscopic theory for their unusual mechanical responses.

\bibliography{refs}

\end{document}